**ORIGINAL RESEARCH**

# Machine learning approach to stock price crash risk

**Abdullah Karasan[1,2]** · **Ozge Sezgin Alp[3]** · **Gerhard-Wilhelm Weber[4,5]**

## Abstract

In this study, we propose a novel machine-learning-based measure for stock price crash risk, utilizing the minimum covariance determinant methodology. Employing this newly introduced dependent variable, we predict stock price crash risk through cross-sectional regression analysis. The findings confirm that the proposed method effectively captures stock price crash risk, with the model demonstrating strong performance in terms of both statistical significance and economic relevance. Furthermore, leveraging a newly developed firm-specific investor sentiment index, the analysis identifies a positive correlation between stock price crash risk and firm-specific investor sentiment. Specifically, higher levels of sentiment are associated with an increased likelihood of stock price crash risk. This relationship remains robust across different firm sizes and when using the detoned version of the firm-specific investor sentiment index, further validating the reliability of the proposed approach.

## 1 Introduction

The relationship between information disclosed to the market and stock price distribution is one of the most important issues attracting researchers and practitioners. In particular, after the 2008 global financial crisis the determinants of negative stock price movements and bad news effect have been at the core of discussion. Most of the studies suggest that stocks are

✉ Abdullah Karasan
   akarasan@umbc.edu

   Ozge Sezgin Alp
   osezgin@baskent.edu.tr

   Gerhard-Wilhelm Weber
   gerhard.weber@put.poznan.pl

1  Adjunct Faculty, University of Maryland, Baltimore County, Baltimore, USA

2  Leveragai, Baltimore, USA

3  Baskent University, Ankara, Turkey

4  Poznan University of Technology, Poznan, Poland

5  IAM, METU, Ankara, Turkey

Springer

more prone to downward movements than upward movements, indicating that stock market returns are asymmetrically distributed in the sense that negative news will cause a larger decline in stock returns than an equal magnitude of good news (Hutton et al., 2009) and (Chen et al., 2001). This asymmetry causes negatively skewed stock returns.

There are four theories that have been offered that seek to explain the reason for negative skewness in stock returns. The first theory is the so-called "leverage effect". First advanced by Black (1976) and then (Christie, 1982), this theory asserts that when company's stock price declines this bad news will increase its leverage. The second theory, the volatility feedback theory, argues that the negative relation between return and volatility as a result of a positive relation between expected risk premiums and volatility (French et al., 1987). In other words, an increase in volatility raises the cost of equity explaining the im- mediate market decline (Pindyck, 1984). A third theory seeking to explain negative skewness is the "stochastic bubbles" theory which states that the negative skewness is due to the popping of asset bubbles, which causes very large market declines, although the probability of this event is low (Banchard & Watson, 1982). During bullish market, investors believe that the demand for stocks will never end which causes market bubbles by leading irrational expectations. This upward trend ends when some investors recognize the unexpected increase and begin selling their stock causes panic selling (Kürüm et al., 2018). The more recent explanation is the discretionary-disclosure theory (Bae et al., 2006). According to this theory, firms prefer to announce good news immediately, but they stockpile negative information. When the accumulated negative information reaches a tipping point, it will cause a large decline. Since concealing bad news about a firm prevents taking timely corrective actions, once the accumulated bad news is released to the market, investors will revise their future expectations and inevitably there will be a sudden decline in prices, which is called as risk (Hutton et al., 2009) and (Kim et al., 2011).

As defined by Hutton et al. (2009) and (Jin & Myers, 2006), crash risk is a large negative market-adjusted re- turn outlier. A number of approaches have been suggested for measuring crash risk literature. (Chen et al., 2001) propose two measures of firm-specific crash risk: (1) The negative co- efficient of skewness of firm-specific returns and (2) The down-to-up volatility. In 2009, (Hutton et al., 2009) introduce the "crash" measure based on the firm-specific weekly returns exceeding 3.09 standard deviations below the mean. Crash risk captures asymmetry in risk that creates an imbalance between risk and return (Kim et al., 2011).

In estimating stock price crash risk, rule-based models are commonplace, that is, if weekly return falls 3.09 standard deviation from the mean, it is labeled as crash risk. Though it sounds appealing and easy-to-apply, rule-based approach comes with cost when (i) The distribution of the stock has changed, and (ii) Researcher lacks domain knowledge. In the presence of these issues, rule-based model can easily result in false negative implying a large number of anomalies are left out of the investigation (Golmohammadi et al., 2014). Alternatively, the model generates false positive indicating a significant number of observations are thought to be anomalies though they are not.

It comes as no surprise that Machine-Learning based models attract great deal of attention as they attack the weakness of the rule-based models and show good predictive performance. The one of the most obvious drawbacks of existing stock price crash measures is their inability to capture changing characteristics of the data as it might not fit well a previously defined rule. This is a problem that can be handled with machine learning models.

Therefore, we seek to estimate stock price crash risk using machine learning- based called Minimum Covariance Determinant (MCD). MCD is a method proposed to detect anomalies in a distribution with elliptically symmetric and unimodal dataset.

Thus, anomalies in stock returns are detected using MCD estimator and this becomes the dependent variable in the logistic panel regression by which we explore the root causes of the crash risk.

The aim of this study is three-fold. Firstly, it attempts to use a machine learning- based crash risk detection method, namely MCD, so that we are able to identify stock price crashes without requiring prior assumptions. Moreover, this method allows us to have a consistent and robust crash risk measure, which improves our understanding towards crash risk determinants. To be more specific, the nexus between stock price crash risk and firm-specific investor sentiment becomes more pronounced and remains robust even after detoning the variable and considering the different firm sizes.

Secondly, a new firm-specific investor sentiment index is proposed. This index is created to understand the link between investor sentiment and stock price crash risk.

Finally, we compare the performance of MCD with other stock price crash risk measures, that is, negative skewness, down-to-up volatility, and crash risk. The findings reveal that the significance level of the estimated variables in our proposed approach is, by and large, higher than those of other models.

These findings contribute to the literature in several ways. First and foremost, aside from previous researches, we embrace machine learning-based model for stock price crash that is adaptable to different state of the financial market. It can also be used to predict future crashes. Failing to predict possible crashes is the main disadvantage of the previous models in the sense that these models tend to fail as they cannot capture the changing characteristics of the data. Secondly, this study exhibits a robust relationship between stock price crash risk and firm-specific investor sentiment. Given the importance of stock price crash drivers, the finding shows how firm-specific investor sentiment is related to the stock price crash. Finally, our findings remain valid after introducing fixed-effect model, controlling for the firm size, and detoning the firm-specific investor sentiment.

The paper proceeds as follows. Section 2 reviews literature on determinants of stock price crash risk. Section 3 describes data and methodology. Section 4 presents the empirical findings and robustness test results and Sect. 5 concludes the paper.

## 2 Literature review

The existing literature extensively explores managerial incentives for hoarding bad news and its implications for stock price crash risk. When managers conceal negative information, the eventual release of accumulated bad news to the market often results in a sudden and significant decline in stock prices, commonly referred to as crash risk. A substantial body of empirical and theoretical research focuses on firm-level determinants of crash risk, examining the factors that drive bad news hoarding behavior through various perspectives. These include the transparency of financial statements (Hutton et al., 2009); (Kim & Zhang, 2014), tax avoidance (Kim et al., 2011), corporate social responsibility (Hunjra et al., 2020); (Kim et al., 2014); (Lee, 2016), accounting conservatism (Kim et al., 2016), CEO overconfidence (Kim et al., 2016), political connections (Luo et al., 2016), directors' and officers' liability insurance (Yuan et al., 2016), corporate philanthropy (Zhang et al., 2016), CEO age (Andreou et al., 2017), internal control (Chen et al., 2017b), and earnings smoothing (Chen et al., 2017a); (Khurana et al., 2018).

In addition to firm-specific indicators, market participants' trading behaviors significantly influence crash risk. For instance, (Callen & Fang, 2013) evaluate the monitoring versus

short- termism perspectives of institutional investors and demonstrate their impact on crash risk. Similarly, (An & Zhang, 2013) identify a negative relationship between crash risk and dedicated investors with strong monitoring roles. On the other hand, (Chen et al., 2017a) show that stock liquidity increases crash risk through transient and non-blockholder investors. Further- more, studies by Huang et al. (2020) and (Vo, 2020) find a positive association between foreign investors and crash risk.

With the advent of behavioral finance, there is growing interest in understanding how irrational investor behavior contributes to stock price dynamics. Pioneering works by Black (1976) and (Long et al., 1990) define irrational investors, or noise traders, as those who make decisions based on random sentiment rather than rational analysis. The uncertainty introduced by noise traders deters rational arbitrageurs from making counter transactions, leading to price deviations that the market fails to correct (Long et al., 1990). Investor sentiment, broadly defined as expectations that deviate from available information about future cash flows and risks (Baker & Wurgler, 2007), reflects investors' optimism or pessimism about stock prices (Baker & Wurgler, 2006). Behavioral finance scholars emphasize investor sentiment as a key driver of stock prices, with studies highlighting its influence on price dynamics (Baker & Wurgler, 2006); (Baker & Wurgler, 2007); (Seok et al., 2019).

While the literature establishes a link between investor sentiment and stock prices, limited research explores managerial behavior in response to investor sentiment. During periods of low sentiment, investors are often pessimistic and prone to undervaluing firms. Managers, in such scenarios, tend to disclose information about future earnings to counteract this pessimism. Conversely, when sentiment is high, managers are less likely to disclose bad news, seeking to sustain the prevailing optimistic environment (Bergman & Roychowdhury, 2008). This behavior aligns with findings that increased investor sentiment amplifies bad news hoarding, thereby raising the likelihood of stock price crashes (Chang et al., 2017); (Cui & Zhang, 2020); (Fu et al., 2020); (Yin & Tian, 2017).

## 3 Data and methodology

The data for this study was collected from the Center for Research in Security Prices (CRSP) and Compustat. The sample includes publicly traded North American companies listed on the New York Stock Exchange (NYSE) and NASDAQ. Weekly stock returns and returns on the value-weighted market index were sourced from CRSP, while balance sheet data and other stock-related information were retrieved from the merged Compustat/CRSP database.

The sample period spans from 1980 to 2020, chosen to ensure a comprehensive histori-cal perspective while utilizing the earliest available data. The dataset under- went rigorous cleaning to ensure robustness and reliability. Specifically:

1. Companies with less than 10% non-zero returns during the sample period were excluded, as their trading activity was deemed insufficient to yield meaningful insights.
2. Firms with fewer than five weekly return observations were also removed to mitigate the risk of distorted results from limited data points.

After these filtering steps, the dataset included 6171 unique firms and 51,903 firm-years, providing a robust sample for the analysis.

To measure firm-specific weekly returns as crash risk indicators, we followed the method-ology outlined in prior literature, including (Chen et al., 2001). This approach isolates firm-specific risk factors, minimizing the influence of broader market movements. Specifi-cally, we employed an expanded market model using ordinary least squares (OLS) estimation to compute residual returns for each firm. The model is represented as follows:

$$r_{j,t} = a_j + \beta 1 j r m, t - 2 + \beta 2 j r m, t - 1$$
$$+ \beta 3 j r m, t + \beta 4 j r m, t + 1 + \beta 5 j r m, t + 2 + \varepsilon j, t, \tag{1}$$

where

- $r_{j,t}$: Weekly return of firm $j$ at time $t$,
- $r_{m,t}$: Value-weighted market return in week $t$,
- $\varepsilon_{j,t}$: Residual return for firm $j$ at time $t$, representing firm-specific factors.

Following the methodology of Kim et al. (2011), the firm-specific weekly returns ($W_{i,t}$) were then computed as:

$$W_{i,t} = \ln(1 + \varepsilon_{i,t}). \tag{2}$$

### 3.1 Crash risk measures

In this part of the study, firstly, theoretical background of MCD along with other stock price crash risk measures are provided. Then, the approach embraced to identify firm- specific sentiment analysis is given.

### 3.1.1 Minimum covariance determinant (MCD)

MCD estimator provides us a robust and consistent method in detecting outliers. In particular, outliers may have huge effect in the multivariate analysis. As summarized by Finch (2012), having outliers in the multivariate analysis can distort correlation coefficient, create multicollinearity issue that makes regression analysis to have biased sample estimates.

Let X be an nxp matrix, if observations are gathered from elliptically symmetric unimodal distribution with $\mu$ and $\Sigma$, then multivariate density function takes the following form (Hubert, 2018):

$$f(x) = \frac{1}{\sqrt{|\Sigma|}} g\big(d^2(x, \mu, \Sigma)\big), \tag{3}$$

where g is strictly decreasing real function, $\mu$ denotes mean vector with $p$ com- ponents, $\Sigma$ is a p × p positive definite covariance matrix, and

$$d^2(x, \mu, \Sigma) = (x - \mu)\Sigma^{-1}(x - \mu), \tag{4}$$

The distance, $d^2$, is known as Mahalanobis squared distance (MSD) that tells us how $X$ is dispersed around the mean of $X$. In this regard, outlier can be thought of as an observation with large Mahalanobis distance from center. However, the conventional.

$\mu$ and $\Sigma$ are not robust to outliers. Thus, robustness is improved by Hardin and Rocke (2002). Based on the half sample h- that is, $h = |(n + p + 1)/2|$, minimum covariance matrix with $\mu^*$ and $\Sigma^*$ is as follows:

$$\mu^* = \frac{1}{n} \sum_{i=1}^{N} x_i, \tag{5}$$

$$\Sigma^* = \frac{1}{n} \Sigma_{i=1}^{N} \big(x_i - \mu^*\big)\big(x_i - \mu^*\big)^T, \tag{6}$$

It is important to note that h does not contain outliers. As the robust MCD is estimated by $h$, outliers do not distort $\mu^*$ and $\Sigma^*$ of MCD. Now, it is time to discuss the way MCD flags the outliers. An observation is labeled as outlier if robust distance, $d^2$, exceeds a cut-off point $c$, which follows $\chi^2_p$ where p represents the degrees of freedom.

In the light of these discussions, the algorithm of MCD can be given as follows:

1. Detect initial robust clustering based on the data,
2. For each cluster, calculate $\mu*$ and $\Sigma*$,
3. Compute MCD, for each observations in the cluster,
4. A new observation with smaller MCD is assigned to the cluster,
5. Half sample h is selected based on smallest MCD and compute $\mu*$
6. and $\Sigma*$ from h,
7. Repeat 2–5 until there is no room for change in h,
8. Detect outlier if $c_p = \sqrt{\chi^2_{p,0.975}}$ is less than $d^2$

The described algorithm for MCD involves several key steps. It begins with an initial clustering of the data and calculates essential cluster properties, such as the mean $\mu^*$ and covariance $\Sigma^*$, for each cluster. The Mahalanobis Distance is then computed for each data point within these clusters, measuring their proximity to the cluster center. Subsequently, data points are assigned to the cluster with the smallest Mahalanobis Distance, ensuring they are grouped with their most similar cluster. A subset of data, known as the half sample, is selected based on these distance measures. The algorithm iteratively refines cluster properties and assignments until the half sample stabilizes. Finally, outliers are detected by comparing Mahalanobis Distances to a threshold (c), typically derived from the chi-squared distribution, and identifying data points with distances less than $d^2$.

In this study, MCD method is applied to detect outliers in stock returns and the result is employed as the dependent variable. Accordingly, if there is a crash risk, the dependent variable takes the value of 1, or 0 otherwise and this dependent variable is called NEGOUTLIER throughout the study.

### 3.1.2 Other crash risk measures

In the literature, three stock price crash measure are widely used, which are NCSKEW, DUVOL, and CRASH. Subsequent to having firm-specific weekly return, NCSKEW and DUVOL can be computed as follows:

$$\text{NSKEW}_{i,t} = \frac{-\left[ n(n-1)^{3/2} \sum w_{i,t}^2 \right]}{\left[ (n-1)(n-2) \left( \sum w_{i,t}^2 \right)^{3/2} \right]} \tag{7}$$

$$\text{DUVOL}_{i,t} = \log \left( \frac{\left[ (n_{u-1}) \sum_{Down} w_{i,t}^2 \right]}{\left[ (n_{d-1}) \sum_{Up} w_{i,t}^2 \right]} \right) \tag{8}$$

Here, $n$ is the number of trading weeks on stock $i$ in year $t$, $n_u$ is the number of up weeks and $n_d$ is the number of down weeks. In a year, weeks with firm-specific returns below the annual mean are called as the down weeks whereas the weeks with firm-specific returns above the annual mean are the up weeks. Thus, the values of NSKEW and DUVOL go in tandem with crash risk.

CRASH measure, on the other hand, is calculated based on the distance from the firm specific weekly returns. That is, CRASH takes the value of 1 if the return is less than 3.2 standard deviations below the mean and 0 otherwise.

## 3.2 Sentiment analysis

Despite some well-known determinants of stock price crash risk, an important aspect of crash risk is thought to be unappreciated, which is firm-specific investor sentiment. It is rather intuitive to say that depending on the positive perception of investor about a company, the stock price might go up or down. That is, if an investor tends to feel optimistic about an individual stock, it is likely that she buys the asset that, in turn, drives the price up or vice versa (Yin & Tian, 2017).

To this respect, in this study, it is aimed to incorporate the firm-specific sentiment to account for the crash risk. That is to say, using variable that have huge impact on investor sentiment towards a company, it is attempted to explore sentiment-stock price crash association. These variables are price-to-earnings ratio (P/E), turnover rate (TURN), equity share (EQS), closed-end fund discount (CEFD), Tobin-q (TOBIN), leverage (LEV), buying and selling volume (BSI).

Detoning, a data preprocessing technique in quantitative finance, is applied to focus exclusively on firm-specific effects by filtering out or neutralizing systematic factors that may obscure the underlying dynamics of financial data.

This process employs statistical methods such as principal component analysis or factor modeling to identify and eliminate influences from factors like market indices or macroeconomic variables. By removing these systematic effects, detoning allows analysts and quants to gain clearer insights into the true, uncorrelated behavior of assets or strategies. This approach is particularly valuable for improving the accuracy of quantitative models, risk assessments, and portfolio optimization in financial analysis.

Previous studies have successfully employed PCA for sentiment analysis and financial modeling. For example, (Baker & Wurgler, 2006) used PCA to construct a market-level sentiment index, demonstrating its ability to aggregate diverse sentiment indicators into a single, coherent measure. This study applies PCA at the firm level, ensuring consistency with established methodologies while addressing firm-specific dynamics. PCA's scalability ensures that the sentiment index remains robust and reliable across different subsets of the data (Jolliffe, 2002).

In order to capture the firm specific sentiment in a proper way, Principal Component Analysis (PCA) are applied using these variables. Then, after obtaining the loadings of the features, we computed the cross-sectional average of components and come up with the following:

$$
\begin{aligned}
SENT_{i,t} = \; & -0.136 P/E_{i,t} + 0.208 TURN_{i,t} \\
& + 0.052 EQS_{i,t} - 0.216 CEFD_{i,t} \\
& + 0.006 TOBIN_{i,t} + 0.17 LEV_{i,t} + 0.048 BSI_{i,t}.
\end{aligned}
\tag{9}
$$

Thus, we derive a time- and firm-specific sentiment index, which is further refined through a process known as detoning. Detoning involves removing the first eigenvector, representing the market-wide component, to isolate the firm-specific effect. This approach is analogous to calculating a beta-adjusted return (Lopez de Prado, 2020). In other words, based on random matrix theory, the first principal component captures the market-wide effect, and excluding it allows us to focus exclusively on the firm-specific sentiment's impact on crash risk in a more filtered and precise manner.

# 4 Empirical findings

This part includes empirical analysis starting from descriptive statistics and correlation analysis. In what follows, the main regression result and robustness checks are provided. In the former analysis, the goodness of our proposed model in detecting the stock price crash is examined by comparing with the existing measures. In the robustness part, the consistency of the result is observed using different estimation technique, varying percentile of firm size, and detoning analysis.

## 4.1 Descriptive statistics and correlation matrix

Table 1 presents the descriptive statistics for all variables used in models. The mean of the most frequently used crash measures NCSKEW and DUVOL are − 0.143 and − 0.179, respectively. The mean value of CRASH is 0.18, indicating 18% of firms experience at least one crash event. This is even lower in NEGOUTLIER crash measure. As expected, the NSCKEW and DUVOL have negative and similar means indicating that these measures concentrate on the sudden stock price drops.

Table 2 presents the correlation matrix for all variables used in models. The correlation coefficient between two main crash risk measures. i.e. NCSKEW and DUVOL is 0.889 and it is significant. The firm specific investor sentiment index is positively significantly correlated with three crash risks measures used in the study. This result points out that as the investor sentiment increase the crash risk tends to rise.

**Table 1** Descriptive statistics

| Variable | Obs | Mean | Std. dev | Min | Max |
|---|---|---|---|---|---|
| NEGOUTLIER | 51,903 | 0.04 | 0.196 | 0 | 1 |
| NCSKEW | 51,903 | − 0.143 | 1.081 | − 7.051 | 5.792 |
| DUVOL | 51,903 | − 0.179 | 1.003 | − 8.313 | 6.59 |
| CRASH | 51,903 | 0.18 | 0.384 | 0 | 1 |
| RET | 45,855 | 0 | 0.002 | − 0.039 | 0.031 |
| SIZE | 45,855 | 6.382 | 2.171 | − 0.34 | 14.466 |
| MTB | 42,212 | 522.992 | 9848.808 | − 962,043.56 | 1,020,718.4 |
| ROA | 45,855 | − 0.027 | 0.378 | − 25.132 | 21.789 |
| SIGMA | 45,855 | 0.014 | 0.009 | 0 | 0.177 |
| SENT | 45,855 | 0 | 0.319 | − 17.58 | 22.076 |
| DTURN | 45,287 | 430.121 | 22,454.041 | − 1,559,962.8 | 2,893,124.8 |
| ACCM | 45,855 | 0.138 | 0.725 | 0 | 70.371 |

**Table 2** Correlation table

| Variables | 1 | 2 | 3 | 4 | 5 | 6 | 7 | 8 | 9 | 10 | 11 | 12 |
|---|---|---|---|---|---|---|---|---|---|---|---|---|
| (1) NEGOUTLIERS | 1.000 | | | | | | | | | | | |
| (2) NCSKEW | 0.273* | 1.000 | | | | | | | | | | |
| (3) DUVOL | 0.217* | 0.889* | 1.000 | | | | | | | | | |
| (4) CRASH | 0.089* | 0.205* | 0.223* | 1.000 | | | | | | | | |
| (5) RET | 0.023* | 0.080* | 0.077* | 0.024* | 1.000 | | | | | | | |
| (6) SIZE | 0.029* | 0.172* | 0.190* | − 0.045* | − 0.027* | 1.000 | | | | | | |
| (7) MTB | 0.009 | 0.013* | 0.015* | − 0.007 | − 0.004 | 0.082* | 1.000 | | | | | |
| (8) ROA | − 0.003 | 0.093* | 0.099* | − 0.018* | 0.049* | 0.243* | 0.015* | 1.000 | | | | |
| (9) SIGMA | 0.011* | − 0.171* | − 0.197* | 0.015* | 0.174* | − 0.417* | − 0.028* | − 0.292* | 1.000 | | | |
| (10) SENT | 0.008 | 0.031* | 0.031* | 0.031* | 0.030* | − 0.107* | − 0.001 | 0.035* | 0.129* | 1.000 | | |
| (11) DVOL | 0.033* | − 0.042* | − 0.040* | − 0.003 | 0.002 | − 0.017* | − 0.001 | − 0.044* | − 0.034* | − 0.177* | 1.000 | |
| (12) ACCM | − 0.003 | − 0.041* | 0.047* | − 0.015* | − 0.026* | − 0.173* | − 0.007 | − 0.212* | 0.131* | 0.013* | 0.011* | 1.000 |

## 4.2 Main results

In order to explore the relationship between stock price crash risk and firm-specific sentiment index, the following the main regression model is constructed. To do that, main independent variables used in the literature are controlled for.

$$
\begin{aligned}
CRASHRISK_{i,t+1} = \ & \alpha_0 + \alpha_1 SENT_{i,t} + \alpha_2 CRASHRISK_{i,t} \\
& + \alpha_3 RET_{i,t} + \alpha_4 SIZE_{i,t} + \alpha_5 MTB_{i,t} \\
& + \alpha_6 ROA_{i,t} + \alpha_7 SIGMA_{i,t} + \alpha_8 TURN_{i,t} \\
& + \alpha_4 ACCM_{i,t} + \Sigma_{t=1}^{T-1} + \Sigma_{k=1}^{K-1} INDUSTRY_k + \varepsilon_{i,t}
\end{aligned}
$$

In this baseline model, the CRASHRISK variable is measured by NEGOUTLIER, CRASH, NCSKEW and DUVOL for year $t + 1$ for $ith$ firm. The primary variable of this section is the firm specific investor sentiment index (SENT) explained in Sect. 2.2. Following the previous literature (Chen et al., 2001); (Hutton et al., 2009); (Kim et al., 2011); (Fu et al., 2020) mean of firm specific return (RET), logarithm of total assets (SIZE), market to book ratio (MTB), return on assets (ROA), standard deviation of firm specific return (SIGMA), detrended turnover ratio (DTURN), discretionary accruals (ACCM) and lagged crash risk measures ($DUVOL_t$ or $NSKEW_t$) at time t are included as control variables.

Detailed definitions of these variables can be found in the Appendix. Having obtained the stock crash risk, it is time to run pooled logistic to explore the relationship with crash risk and crash risk determinants.

While conducting the regressions to control for industry-invariant and time-invariant variables, the time and industry fixed effects are also incorporated into the model. In addition, standard errors are corrected for cross-sectional heteroskedasticity.

The result is summarized in Table 3. The first Model (NEGOUTLIER) is our proposed Model. In Models 2, 3, and 4, we utilize CRASH, NCSKEW, and DUVOL measure to see observe the consistency of the results.

As is seen from the Table 3, for all crash risk measures (NSKEW, DUVOL, CRASH and NEGOUTLIER), the coefficients of firm specific investor sentiment index are positive and statistically significant at 1% level. In the literature, in times of high sentiment, under pressure of optimistic expectations, managers tend to accelerate good news but withhold bad news to maintain the positive environment (Bergman & Roychowdhury, 2008). Thus, the result suggests a positive relationship between sentiment and crash risk (Cui & Zhang, 2020); (Fu et al., 2020); (Yin & Tian, 2017). Presence of high correlation between crash risk and firm specific investor sentiment index across alternative measures verifies the robustness.

Prior findings state that in the presence of negative skewness, stock price crash risk tends to increase (Chen et al., 2001); (Jin & Myers, 2006); (Kim & Zhang, 2014). Therefore, we control for the weekly return skewness. NCSKEW exerts similar statistically significant result across all measures. However, the magnitude of its impact is higher in the first Model. Moreover, SIZE, RET, MTB, and SIGMA have all positive and significant result. More elaborately, SIZE is highly significant and financially important variable with its high magnitude estimated coefficient. This observation implies high likelihood of negative skewness in large-cap companies.

On the contrary, MTB has low significance and trivial financial impact on stock price crash risk. Another interesting observation would be that MTB is only significant in only our proposed Model presented in the first column. As expected, past return is also positive and highly significant implying that high past returns are predicted to have more negative skewness and the magnitude is even higher in our Model.

**Table 3** Main model results

| | NEGOUTLIER$_{t+1}$ | CRASH$_{t+1}$ | NCSKEW$_{t+1}$ | DUVOL$_{t+1}$ |
|---|---|---|---|---|
| NCSKEW$_t$ | 0.163*** | 0.0333** | | 0.0480*** |
| | (0.0293) | (0.0150) | | (0.00614) |
| RET$_t$ | 0.916*** | 0.402*** | 0.757*** | 0.667*** |
| | (13.96) | (7.645) | (3.896) | (3.344) |
| SIZE$_t$ | 0.131*** | − 0.0530*** | 0.0654*** | 0.0644*** |
| | (0.0145) | (0.00725) | (0.00304) | (0.00277) |
| MTB$_t$ | 2.56e−06* | − 8.99e−07 | − 1.05e−07 | 1.11e−08 |
| | (1.50e−06) | (8.41e−07) | (3.42e−07) | (3.04e−07) |
| ROA$_t$ | − 0.0969* | − 0.0217 | 0.0705*** | 0.0635*** |
| | (0.0542) | (0.0316) | (0.0198) | (0.0198) |
| SIGMA$_t$ | 12.00*** | 2.480 | − 13.39*** | − 13.20*** |
| | (3.122) | (1.789) | (0.885) | (0.790) |
| SENT$_t$ | 0.106** | 0.209*** | 0.141*** | 0.134*** |
| | (0.0442) | (0.0613) | (0.0369) | (0.0339) |
| DTURN$_t$ | 3.91e−06 | 1.81e−07 | − 1.70e−06** | − 1.49e−06** |
| | (2.43e−06) | (5.42e−07) | (8.19e−07) | (7.38e−07) |
| ACCM$_t$ | − 0.0213 | 0.00884 | 0.00280 | − 0.00140 |
| | (0.0503) | (0.0139) | (0.00734) | (0.00693) |
| DUVOL$_t$ | | | 0.0673*** | |
| | | | (0.00765) | |
| Constant | − 3.421*** | − 1.321*** | − 0.316*** | − 0.392*** |
| R-squared | | | 0.075 | 0.094 |
| Industry fixed effect | Yes | Yes | Yes | Yes |
| Year fixed effect | Yes | Yes | Yes | Yes |

Dependent variables are the NEGOUTLIER, CRASH, NCSKEW, and DUVOL at time $t + 1$
Standard errors are reported in parentheses. *$p < 0.1$, **$p < 0.05$, ***$p < 0.01$

It is found out that the stock performance is negatively related with stock price crash risk. It turns out our Model confirms the finding in the literature in that the ROA has negative and significant (Hutton et al., 2009). In Model 2 of Table 3, ROA does not have any explanatory power and, in Model 3 and 4 of Table 3, ROA has positive relation with stock price crash risk measures. In general, companies with higher ROAs tend to be more profitable and financially stable.

Higher profitability can signal to investors that a company is better equipped to economic downturns and unexpected shocks. As a result, companies with higher ROAs may be perceived as having lower stock price crash risk.

Another striking observation is about the standard deviation of weekly return. SIGMA shows a very significant and large financial impact on all Models except for CRASH. As high volatility results from high trading volume, stocks with high volatility tends to have higher likelihood of stock price crashes. However, it is not the case for the Models 3 and 4 of the same Table in which we have negative and significant estimated SIGMA coefficient.

As shown above, the results show that Model 1 of Table 3 does not only have many significant and financial important variables but also the sign and magnitude of these variables are consistent with prior literature.

## 4.3 Robustness checks

In this section, we conduct a series of robustness checks to validate the relationship between the firm-specific investor sentiment index and stock price crash risk.

These checks address potential econometric challenges, including endogeneity, heterogeneity across firm sizes, and the influence of market-wide factors, ensuring the reliability and robustness of the results. A multi-faceted approach is adopted, encompassing firm fixed effects, size-specific analyses, detoning procedures, dynamic panel estimations, and machine learning classification models.

First, recognizing that the impact of investor sentiment varies across firms of different sizes, we analyze the relationship between stock price crash risk and explanatory variables within size-based firm categories. Previous studies suggest that smaller firms are more vulnerable to speculative activity during periods of heightened sentiment due to their lower visibility among arbitrageurs and slower price adjustments. Conversely, larger firms, where mispricing is corrected more efficiently, exhibit weaker sentiment effects.

By categorizing firms into size quartiles, this analysis provides a nuanced understanding of how sentiment-crash risk dynamics vary across firm sizes, shedding light on the heterogeneity of investor sentiment effects. Second, we examine the relationship between stock price crash risk and explanatory variables across different firm sizes. Recognizing that the impact of investor sentiment and other factors may vary depending on firm characteristics, this analysis provides insights into size-related heterogeneity in crash risk dynamics.

Second, a detoning procedure is applied to ensure that the firm-specific investor sentiment measure captures idiosyncratic effects rather than market-wide movements. This involves removing the first principal component from the sentiment index, effectively filtering out systematic market influences.

By focusing on firm-specific sentiment, the analysis avoids potential biases introduced by broader market dynamics. Regression models are subsequently re-estimated using the detoned sentiment index to examine whether the findings remain consistent under this adjustment.

Finally, machine learning classification models are utilized to further assess the relationship between independent variables and crash risk. Decision Tree, Random Forest, and XGBoost models are employed to evaluate the predictive accuracy of the independent variables, as measured by Root Mean Square Error (RMSE). Feature importance analysis is conducted to identify the most influential predictors of stock price crash risk, providing additional validation for the robustness of the results.

Together, these methods provide a comprehensive framework to assess the robust- ness of the relationship between firm-specific investor sentiment and stock price crash risk, offering deeper insights into both static and dynamic aspects of this association.

### 4.3.1 Controlling for the investor sentiment size proxy

The impact of investor sentiment on stock returns is significantly influenced by firm size. Prior literature, such as studies by Kim and Ha (2010); Qadan & Aharon, 2019), and (Seok et al., 2019), has extensively examined how investor sentiment affects small, medium, and large firms. For instance, (Qadan & Aharon, 2019) argued that market participants tend to overvalue small stocks relative to large stocks, with sentiment explaining the return differences. (Seok et al., 2019) and (Kim & Ha, 2010) further demonstrated that firm size alters the relationship between sentiment and stock price movements. Small firms, which are often undervalued and less scrutinized by arbitrageurs, be- come more attractive to speculators during periods of high sentiment (Baker & Wurgler, 2007). This lack of arbitrage activity for small firms makes price adjustments slower, increasing the likelihood of stock price crashes. Conversely, large firms, where mispricing is corrected more swiftly, exhibit a weaker sentiment effect (Seok et al., 2019).

To investigate this relationship further, firms were categorized into size-based groups, and the analysis focused on the relationship between investor sentiment and stock price crash risk.

Table 4 presents descriptive statistics for firms across different size quintiles, indicating that the mean firm-specific investor sentiment is higher for smaller firms. This suggests that smaller firms are more susceptible to sentiment- driven mispricing. To test whether firm size alters the sentiment-crash risk relation- ship, dynamic panel estimation is used.

### 4.3.2 Dynamic panel estimation

Dynamic panel estimation approach addresses potential endogeneity issues arising from lagged dependent variables and unobserved firm-specific effects. To do that Arellano–Bond estimator (Arellano & Bond, 1991) is employed. By using lagged values of explanatory variables as instruments, the Arellano-Bond GMM ensures consistent and efficient parameter estimation.

Specifically, it is designed for situations where the number of time periods (T) is small relative to the number of entities (N), such as firms in this study. The Arellano-Bond GMM is employed here to model the dynamic effects of investor sentiment and other covariates on crash risk (NEGOUTLIER). This method helps account for reverse causality and omitted variable bias, making the results more robust. The model is estimated for each size group using the NEGOUTLIER variable as the crash risk measure. Table 5 summarizes these results.

Table 5 displays the regression results for large and small firms. Consistent with prior literature, the coefficient for firm-specific sentiment ($SENT_t$) is positive and statistically significant for small firms, indicating that heightened sentiment increases crash risk for these firms. In contrast, the sentiment coefficient for large firms is negative and statistically significant in the second column of 5, reflecting the weaker sentiment effects due to efficient price corrections.

The analysis reveals several key observations. First, the lagged NEGOUTLIER coefficients (L1.NEGOUTLIER and L2.NEGOUTLIER) are significant for small firms, indicating that past crash risk plays an important role in shaping future crash risk dynamics for small firms. Additionally, variables such as SIZE and SIGMA exhibit differing impacts depending on firm size, emphasizing the necessity of accounting for size heterogeneity when examining sentiment effects. Finally, the relationship between investor sentiment and crash risk is stronger for smaller firms, supporting the notion that these firms attract speculative activity during periods of heightened sentiment (Baker & Wurgler, 2007).

As suggested by Billio et al. (2010); Kim & Jeong, 2005), and (Zheng et al., 2012), while creating the investor sentiment index, if the first component of PCA is dropped, the systematic effect can be filtered out. In doing so, it is anticipated to have more firm-oriented sentiment analysis, which provides us more robust result. This analysis is referred to as detoning analysis.

**Table 4** Descriptive statistics (Based on Size)

| | Mean | 4th Quartile (Large Firms) Std. Dev | Mean | 3rd Quartile Std. Dev | Mean | 2nd Quartile (Small Firms) Std. Dev | Mean | 1st Quartile Std. Dev |
|---|---|---|---|---|---|---|---|---|
| NEGOUTLIER | 0.0439 | 0.2049 | 0.0404 | 0.1969 | 0.0453 | 0.2080 | 0.0300 | 0.1708 |
| CRASH | 0.1548 | 0.3618 | 0.1836 | 0.3872 | 0.1844 | 0.3878 | 0.1977 | 0.3982 |
| NCSKEW | 0.0258 | 0.8904 | − 0.0316 | 0.9766 | − 0.1126 | 1.073 | − 0.452 | 1.278 |
| DUVOL | 0.0063 | 0.8535 | − 0.0655 | 0.9231 | − 0.1653 | 0.9931 | − 0.4866 | 1.144 |
| MINRET | 0.0258 | 0.0186 | 0.0322 | 0.0212 | 0.0399 | 0.0247 | 0.0483 | 0.0296 |
| RET | − 0.0001 | 0.0014 | − 0.0000 | 0.0018 | − 0.0001 | 0.0022 | − 0.0000 | 0.0030 |
| SIZE$t$ | 92.164 | 11.441 | 7.019 | 0.4397 | 5.555 | 0.4264 | 3.644 | 0.8352 |
| MTB$t$ | 1734.46 | 1989 | 223.96 | 2493.4 | 129.61 | 3304 | 118.71 | 3,136.2 |
| ROA$t$ | 0.0452 | 0.0766 | 0.0345 | 0.1433 | − 0.0155 | 0.3590 | − 0.2139 | 14.506 |
| SIGMA$t$ | 0.0101 | 0.0059 | 0.0124 | 0.0068 | 0.0155 | 0.0086 | 0.0203 | 0.0126 |
| SENT$t$ | − 0.0489 | 0.2400 | − 0.0113 | 0.2733 | 0.0203 | 0.2423 | 0.0418 | 0.6298 |
| DTURN$t$ | 259.776 | 1957 | 153.58 | 8031.7 | 52.62 | 8582 | 1279.6 | 3.9024 |
| ACCM$t$ | 0.0348 | 0.0699 | 0.0630 | 0.4251 | 0.1064 | 0.5021 | 0.3750 | 2.119 |

**Table 5** Arellano-Bond GMM

| | NEGOUTLIER$_{t+1}$ (Large Firms) | NEGOUTLIER$_{t+1}$ | NEGOUTLIER$_{t+1}$ | NEGOUTLIER$_{t+1}$ (Small Firms) |
|---|---|---|---|---|
| $L_1$.NEGOUTLIER | − 0.0252 | − 0.5036 | − 1.2778*** | 0.9022** |
| | (0.0560) | (0.9420) | (0.2232) | (0.3408) |
| $L_2$.NEGOUTLIER | 0.0833 | 0.1021 | − 0.6289*** | 0.1934 |
| | (0.1571) | (0.1453) | (0.1147) | (0.1170) |
| RET$_t$ | − 0.0056* | − 0.0562 | − 0.0421 | 0.1244** |
| | (0.0028) | (0.0763) | (0.0716) | (0.0449) |
| SIZE$_t$ | 0.0149 | 0.0222** | 0.0599** | − 0.0233 |
| | (0.0116) | (0.0070) | (0.0221) | (0.0168) |
| MTB$_t$ | − 0.0071 | 0.0145 | − 0.0161 | − 0.0026 |
| | (0.0930) | (0.0624) | (0.0644) | (0.0014) |
| ROA$_t$ | 0.0005 | 0.7572 | 1.1120 | 0.8455** |
| | (0.0006) | (1.0474) | (1.2755) | (0.3026) |
| SIGMA$_t$ | − 0.0050 | 0.3610 | 0.1405 | 0.3772** |
| | (0.0083) | (0.4580) | (0.5207) | (0.1347) |
| SENT$_t$ | − 0.0048 | − 0.2313* | 0.2650** | 0.3883** |
| | (0.0078) | (0.0973) | (0.0879) | (0.1470) |
| DTURN$_t$ | − 0.0000 | 0.0000 | − 0.0001** | 0.0000 |
| | (0.0000) | (0.0001) | (0.0000) | (0.0000) |
| NCSKEW$_t$ | 0.1639*** | 0.7488** | 0.3649*** | 0.3628*** |
| | (0.0486) | (0.2327) | (0.0483) | (0.0700) |

**Table 5** (continued)

| | NEGOUTLIER$_{t+1}$ (Large Firms) | NEGOUTLIER$_{t+1}$ | NEGOUTLIER$_{t+1}$ | NEGOUTLIER$_{t+1}$ (Small Firms) |
|---|---|---|---|---|
| Constant | 0.0008 | − 0.0164 | 0.0207*** | − 0.0257* |
| | (0.0024) | (0.0298) | (0.0049) | (0.0114) |

Dependent variable is NEGOUTLIER with different firm sizes. Standard errors are reported in parentheses. $*p < 0.1$, $**p < 0.05$, $***p < 0.01$

**Table 6** Detoning Analysis

| | NEGOUTLIER$_{t+1}$ (Large Firms) | NEGOUTLIER$_{t+1}$ | NEGOUTLIER$_{t+1}$ | NEGOUTLIER$_{t+1}$ (Small Firms) |
|---|---|---|---|---|
| NCSKEW$_t$ | 0.0865 | 0.230*** | 0.0365 | 0.190*** |
| | (0.0610) | (0.0558) | (0.0630) | (0.0699) |
| RET$_t$ | 86.61*** | 116.3*** | 69.01** | 77.86** |
| | (24.78) | (25.62) | (32.76) | (39.49) |
| SIZE$_t$ | 0.451*** | 0.257** | 0.292*** | 0.0973* |
| | (0.0896) | (0.102) | (0.106) | (0.0545) |
| MTB$_t$ | 3.06e−05*** | 3.52e−06 | 2.00e−06 | 2.87e−06* |
| | (1.15e−05) | (4.16e−06) | (2.24e−05) | (1.63e−06) |
| ROA$_t$ | − 0.140* | − 0.0373 | 0.458* | − 0.487 |
| | (0.0809) | (0.0929) | (0.239) | (0.730) |
| SIGMA$_t$ | 0.631 | 4.254 | 41.02*** | 33.28*** |
| | (5.662) | (6.631) | (7.418) | (9.157) |
| SENT$_t$ | 0.594*** | 0.0604 | − 0.344* | 0.202 |
| | (0.199) | (0.0506) | (0.181) | (0.301) |
| DTURN$_t$ | 1.29e−06** | 3.40e−05*** | 4.40e−05*** | 1.09e−05** |
| | (5.37e−07) | (6.67e−06) | (9.49e−06) | (4.58e−06) |
| ACCM$_t$ | − 0.0367 | − 0.0368 | 0.136** | − 1.798* |
| | (0.0548) | (0.137) | (0.0684) | (0.988) |
| Constant | − 4.475*** | − 3.597*** | − 4.775*** | − 3.603*** |
| | (0.794) | (0.886) | (1.152) | (1.329) |
| Industry FE | Yes | Yes | Yes | Yes |
| Year FE | Yes | Yes | Yes | Yes |

Dependent variable is NEGOUTLIER with different firms sizes. Standard errors are reported in parentheses. *$p < 0.1$, **$p < 0.05$, ***$p < 0.01$

The result of the detoning analysis is given in the Table 6. The result reveals that sentiment variable keeps its statistical significance for the large (first quantiles) and medium-sized (third quantiles) firms. To be more specific, for small firms, sentiment analysis is positively correlated with NEGOUTLIER at 1% significance level. How- ever, the direction of association is reversed when it comes to firms in the second quantile corresponding to third column of Table 6.

The result suggests that, at 10% significance level, the relationship between sentiment variable and NEGOUTLIER is statistically significant, that is, one-point increase in sentiment variable leads to decrease in NEGOUTLIER by 0.0344 point. This observation confirms that small firms attract speculators attention and this in turn may lead to stock price crash. This association is not that strong for other firm sizes, because big firms is liquid and attract arbitragers and therefore price adjustment happens quickly.

As a final robustness check, we use machine learning classification models to evaluate the impact of independent variables on crash risk, specifically NEGOUT- LIER. The results presented in Table 7 highlight the predictive performance of three machine learning models: Decision Tree, Random Forest, and XGBoost, based on their Root Mean Square Error (RMSE) values. RMSE, a widely used metric for assessing prediction accuracy, measures the average magnitude of errors in predictions, with lower values indicating better predictive performance.

**Table 7** Comparison of RMSE values for machine learning models

| Model | RMSE |
|---|---|
| Decision tree | 0.62 |
| Random forest | 0.64 |
| XGBoost | 0.68 |

The Decision Tree model achieves the lowest RMSE (0.62), followed by Random Forest (0.64) and XGBoost (0.68). These results suggest that the Decision Tree model outperforms the other methods on this dataset in terms of minimizing prediction error.

We then use the XGBoost model to analyze the impact of variables on crash risk by plotting their feature importance. Figure 1 demonstrates that the turnover ratio (dturn) and size (log_size) are, respectively, the most and least impactful variables influencing the dependent variable.

Figure 1 exhibits that turnover ratio (dturn) and size (log_size) are the most and least effective variables respectively on the dependent variable.

The SHAP value given in the Fig. 2 analysis highlights the key drivers of stock price crash risk. Turnover (dturn) emerges as a significant factor, with higher values substantially increasing crash risk. Similarly, firm-specific sentiment positively correlates with crash risk, indicating that optimistic sentiment contributes to risk, aligning with managerial tendencies to withhold bad news. Negative skewness (NCSKEW) also shows a strong positive impact, consistent with established crash risk theories. In contrast, return on assets (RoA) reduces crash risk, acting as a stabilizing factor, while firm size has minimal influence. Overall, the analysis underscores turnover, sentiment, and negative skewness as critical predictors of crash risk, supporting the model's robustness.

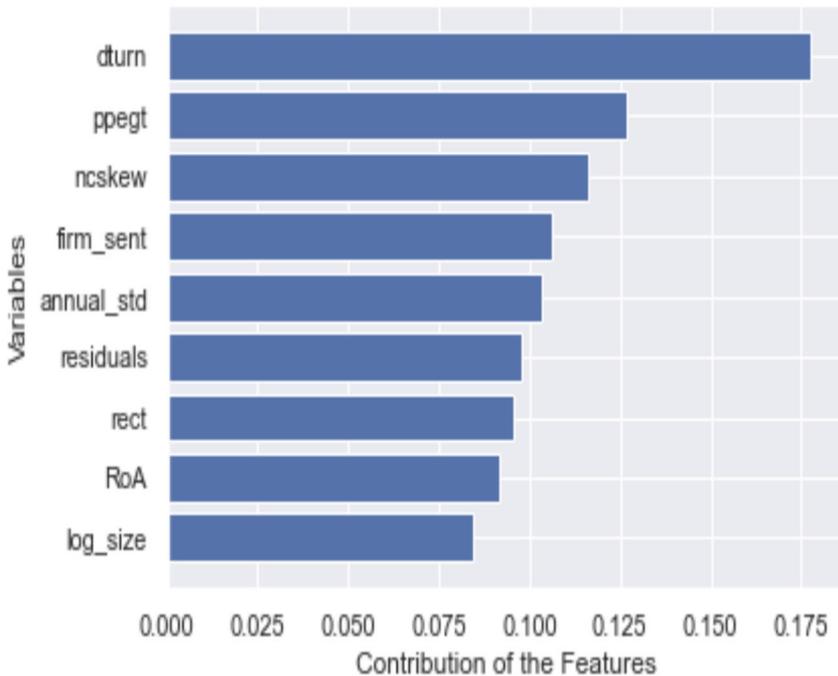

**Fig. 1** Feature importance

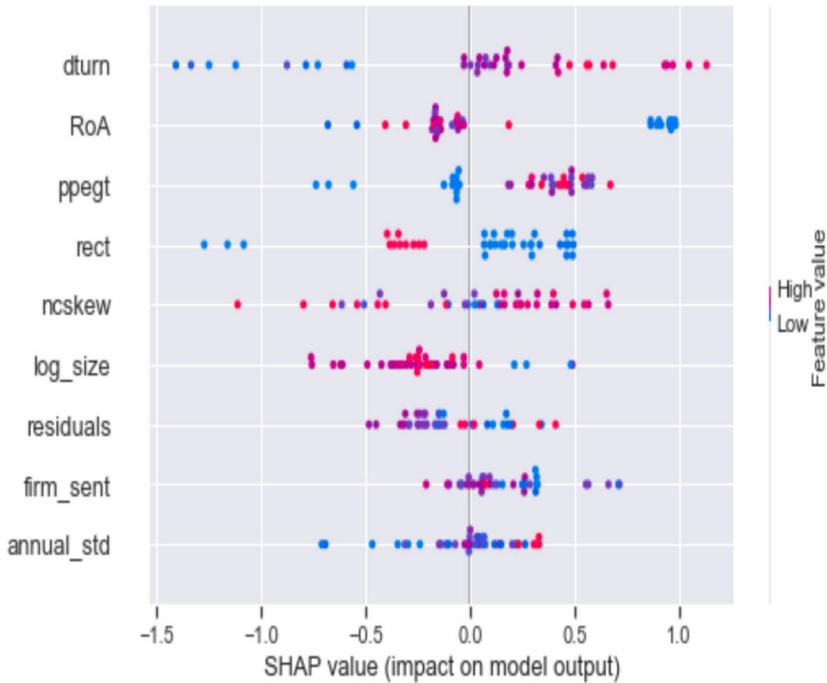

**Fig. 2** Feature importance

## 5 Conclusion

In this study, we present a novel machine-learning-based stock price crash risk measure utilizing the Minimum Covariance Determinant (MCD) method. This approach detects negative anomalies in market-adjusted firm-specific stock price returns and defines them as indicators of stock price crash risk. The performance of the MCD-based crash risk measure is found to align with well-established measures such as negative skewness, down-to-up volatility, and traditional crash risk indicators.

Additionally, this study examines the impact of firm-specific investor sentiment on stock price crash risk. To effectively capture firm-specific sentiment, we apply Principal Component Analysis (PCA) to variables likely to influence sentiment. The findings reveal a positive relationship between investor sentiment and crash risk, indicating that during periods of heightened sentiment, driven by optimistic expectations, managers are more inclined to withhold bad news. This accumulation of with- held bad news eventually results in significant declines in stock prices. The use of a firm fixed-effects model further substantiates the relationship between firm-specific investor sentiment and stock price crash risk, demonstrating a stronger financial magnitude compared to other measures. Moreover, this relationship remains robust after controlling for firm size and applying detoning effects.

The proposed machine-learning-based model offers several advantages over traditional approaches. First, it is data-dependent, enhancing its adaptability to diverse datasets. Second, it establishes a robust, strong, and statistically significant relationship between firm-specific investor sentiment and stock price crash risk. Lastly, the model facilitates forecasting, providing insights into potential future stock price crashes.

Despite its strengths, this study has certain limitations. One limitation is the exclusion of additional behavioral variables, such as CEO confidence and accounting conservatism, which could further enrich the analysis. Another limitation is the absence of an optimized approach to identify the lead-lag relationships between crash risk and the independent variables. These aspects present opportunities for future re- search to build upon the findings of this study.

## Appendix

*NEGOUTLIER:* The machine learning based stock crash risk measure which uses Minimum Covariance Determinant (MCD) estimator to detect anomalies in the distribution of firm-specific weekly returns. The negative outliers are classified as crash events once it takes the value of 1.

*CRASH:* The crash risk measure that takes value of 1 if the firm specific weekly returns exceed 3.2 standard deviations below the average and zero otherwise.

*NCSKEW:* The negative coefficient of skewness of firm-specific weekly returns in a year which is the negative of third moment of firm-specific weekly returns, divided by the cubed standard deviation.

*DUVOL:* The down-to-up volatility measure of firm specific return in a year which is the logarithm of the standard deviation on down weeks divided by the standard deviation on up weeks.

*RET:* The average of firm-specific weekly return in a year.

*SIZE:* The natural logarithm of total assets in a year.

*MTB:* The ratio of market capitalization of equity to the book value of equity in a year.

*ROA:* The returns on asset in a year which is the ratio of net income to total assets.

*SIGMA:* The standard deviation of firm-specific weekly return in a year.

*SENT:* The firm-specific investor sentiment measure obtained by Principal Component Analysis (PCA).

*DTURN:* The average monthly turnover ratio in year t minus the average monthly turnover ratio in year t-1. The turnover ratio is the monthly trading volume divided by the total number of shares outstanding.

*ACCM:* The absolute value of discretionary accruals in a year. The discretionary accruals are estimated from the modified Jones model.

## Declarations

of funding, potential conflicts of interest (financial or non-financial), informed consent if the research involved human participants, and a statement on welfare of animals if the research involved animals.